\def\year{2019}\relax
\documentclass[letterpaper]{article} 
\usepackage{aaai19}  
\usepackage{times}  
\usepackage{helvet}  
\usepackage{courier}  
\usepackage{url}  
\usepackage{graphicx}  
\usepackage{wasysym} 
\usepackage{verbatim} 
\usepackage{amsfonts} 
\usepackage{amssymb} 
\usepackage{amsthm} 
\newcommand{\intension}[2]{[\![ #1 ]\!]_{ #2}}

\usepackage[shortlabels]{enumitem} 

\newtheorem{theorem}{Theorem}[section]
\newtheorem{example}[theorem]{Example}
\newcommand{\ag}{{\mathit{ag}}}
\newcommand{\Ag}{{\mathit{Ag}}}
\newcommand{\yes}{{$\mathsf{yes}$}}
\newcommand{\no}{{$\mathsf{no}$}}
\newcommand{\U}{{\mathcal U}}
\newcommand{\V}{{\mathcal V}}
\newcommand{\R}{{\mathcal R}}
\newcommand{\F}{{\mathcal F}}

\newcommand{\E}{{\mathcal E}}
\newcommand{\K}{{\mathcal K}}
\newcommand{\A}{{\mathcal A}}
\newcommand{\G}{{\mathcal G}}
\newcommand{\db}{\mathit{db}}
\newcommand{\gb}{\mathit{gb}}
\newcommand{\mb}{\mathit{mb}}
\newcommand{\commentout}[1]{}
\newcommand{\wbox}{\mbox{$\sqcap$\llap{$\sqcup$}}}

\begin{document}
\title{Blameworthiness in Multi-Agent Settings}
\author{Meir Friedenberg \\
	Department of Computer Science\\
	Cornell University\\
	meir@cs.cornell.edu 
	\And 
	Joseph Y. Halpern\\
	Department of Computer Science\\
	Cornell University\\
	halpern@cs.cornell.edu\\
}
\maketitle
\begin{abstract}
	We provide a formal definition of blameworthiness in settings
	where multiple agents can collaborate to avoid a negative
	outcome.  We first provide a method for ascribing
	blameworthiness to groups relative to an epistemic state (a
	distribution over causal models that describe how the outcome
	might arise).  We then show how we can go from an ascription
	of blameworthiness for groups to an ascription of
	blameworthiness for individuals using a standard notion from
	cooperative game theory, the \emph{Shapley value}. We believe
	that getting a good notion of blameworthiness in a group
	setting will be critical for designing autonomous agents that
	behave in a moral manner. 
\end{abstract}

\section{Introduction}
As we move towards an era where autonomous systems are ubiquitous, being able to
reason formally about moral responsibility will become more and more
critical.  Such reasoning will be necessary not only for legal
ascription of responsibility, but also in order to design systems that
behave in a moral manner in the first place.  Unfortunately, though,
pinning down the many notions related to moral responsibility has been
notoriously difficult to do, even informally.
In this work, we lay foundations on which these problems can be solved.

Halpern and Kleiman-Weiner \shortcite{HKW18} (HK from now on) made important headway on
this work by providing a definition of blameworthiness based on a
causal framework.  An epistemic state, that is, a distribution over
causal models, can be used to capture an agent's beliefs about the
effects that actions may have.  Given an epistemic state, they then
provide a definition of blameworthiness for an outcome by taking into
account both whether the agent believed they could affect the
likelihood of the outcome and the cost they believed would be
necessary to do so. 

While their definition seems compelling in single-agent settings, as
HK themselves observe, the definition does not capture
blameworthiness in multi-agent settings where, if the group could
coordinate their actions,
they could easily bring about a different
outcome (see Section~\ref{sec:groupblame} for more discussion of this issue). 
Being able to analyze
blameworthiness in multi-agent scenarios seems critical in practice;
if an accident is brought about due to the actions of multiple agents,
we would like to understand to what extent each one is blameworthy.

We tackle that problem here by defining a
notion of group blameworthiness in multi-agent scenarios.
Our notion can be viewed as a generalization of the single-agent
notion of blameworthiness defined by HK.
However, as we shall see, subtleties arise when considering groups.
Once we have a way of ascribing blameworthiness to a group,
we show that a standard notion from cooperative game
theory, the \emph{Shapley value} \cite{Shapley1953}, can be used to
apportion blame to individual members of the group, 
and is in fact the only way to do so that satisfies a number of
desirable properties. 

The rest of this paper is organized as follows.  In
Section~\ref{sec:causal} we review the basic causal framework used.
Section~\ref{sec:blame} constitutes the core of the work; in it, we
review the Halpern and Kleiman-Weiner definition of blameworthiness in
the single-agent setting, define group blameworthiness, show how to
apportion it to individual agents, and demonstrate how the
definition plays out in an illustrative example.  In
Section~\ref{sec:related} we review related work and in
Section~\ref{sec:conclusion} we conclude. 

\section{Causal Models}\label{sec:causal}

At the heart of our definitions is the causal model framework of
Halpern and Pearl \shortcite{HP01b}.
We briefly review the relevant material here.

A causal situation is characterized by a set of variables and their
values.  A set of \textit{structural equations} describes the effects
that the different variables have on each other.  Among the variables
we distinguish between \textit{exogenous variables} (those variables
whose values are determined by factors not modeled) and \textit
{endogenous variables} (those variables whose values are determined by
factors in the model). 

A causal model $M = (\mathcal{S},\F)$ consists of a signature
$\mathcal{S}$ and a set $\F$ of structural equations.  A signature
$\mathcal{S} = (\U,\V,\R)$ is in turn composed of a (finite but
nonempty) set $\U$ of exogenous variables, a (finite but nonempty)
set $\V$ of endogenous variables, and a range function $\R$ that maps
each variable in $\U \cup \V$ to a set of values it can take on.
$\F$ associates with each endogenous variable $X \in \V$ a
function denoted $F_X$ such that $F_X: (\times_{U \in \U} \R(U))
\times (\times_{Y \in \V - \{X\}} \R(Y)) \rightarrow \R(X)$;
that is, $F_X$ determines the value of $X$,
given the values of all the other variables in $\U \cup \V$.

We assume that there is a
special subset $\A$ of the endogenous variables
known as the \textit{action variables}.
Since we consider scenarios with multiple agents,
we need to be able to identify which agent each
action is associated with.  Thus, given a set $G$ of agents, we
augment the signature of the causal model to be $\mathcal{S} =
(\U,\V,\R,\G)$, where $\G:\A \rightarrow G$ associates an agent with
each action variable in $\A$.
The range of a variable $A \in \A$ is simply
the set of actions available to the agent $\G(A)$.  So, for
instance, if there are five members of a committee and each of them
can vote \yes\ or \no\ then there will be action variables $A_1$
through $A_5$ with $\G(A_i) = i$ and $\R(A_i) = \{ $\yes$,$\no$ \}$
for all $i$.
We restrict here to scenarios where there is one action per
agent; in future work, we hope to consider blameworthiness
in planning scenarios, where agents may take multiple
actions sequentially. 

In causal models, we can reason about interventions.  Specifically, we
have formulas of the form $[A \gets a]\varphi$, which can be read ``if
action $a$ were performed, then the outcome would be $\varphi$'', where
an \emph{outcome} is a Boolean combination of primitive events of the
form $X=x$.  We give semantics to such formulas in a \emph{causal
setting} $(M,\vec{u})$ consisting of a causal  model $M$ and a \emph{context}
$\vec{u}$, an assignment of values to all the exogenous variables.
We do not need the details of the semantics in what follows; they can
be found in \cite{Hal48,HP01b}.  

\section{Blameworthiness}\label{sec:blame}

With this background, we now turn to the question of how blameworthy
an agent $\ag$ is for an outcome $\varphi$.  We begin by reviewing
the HK definition, and then propose a way of dealing with settings
that allow coordinated group actions.
One caveat: as noted by HK, words like ``blame'' have a wide
variety of nuanced 
meanings in natural language. While we think that the notion that we are trying
to capture (which is essentially the same as the notion that HK
tried to capture) is useful,  it corresponds at best to only one 
way that the word ``blame'' is used by people.

\subsection{Blameworthiness in a single-agent setting}

HK identified two factors that
play a role in determining blameworthiness: $\ag$'s beliefs about his
ability to affect $\varphi$ and $\ag$'s beliefs about the cost necessary
to affect $\varphi$.  Here we present a slightly simplified version of
their formalization of these notions. 

An agent $\ag$ has an \textit{epistemic state} $\mathcal{E} =
(Pr,\mathcal{K})$ relative to which his or her blameworthiness is
determined.  $\mathcal{K}$ is the set of all causal settings that $\ag$
considers possible, and $Pr$ is a
probability on $\K$.%
\footnote{In HK's original formalization, an epistemic
state also contained a utility function. 
This is not necessary for our purposes, so to simplify matters we
leave it out.}
Given $\mathcal{E} = (\Pr,\K)$, two actions $a$ and $a'$, and an outcome
$\varphi$, we can define $\delta^\E_{a,a',\varphi}$ to be how much more
likely $\varphi$ was to occur if the agent performed action $a$ than
if he performed $a'$.  Let $\intension{ \psi}{\mathcal{K}}$
denote the set of all settings in $\mathcal{K}$ where $\psi$ is true, so
that $Pr(\intension{ [A=a] \varphi }{\mathcal{K}})$ is the
probability that $\ag$ ascribes to outcome $\varphi$ occurring given
that action $a$ is taken.  Then $\delta^\E_{a,a',\varphi} = \max
(0,Pr(\intension{[A=a] \varphi}{\mathcal{K}}) -
Pr(\intension{ [A=a'] \varphi}{\mathcal{K}}))$.
Thus, $\delta^\E_{a,a',\varphi}$ is $0$ if performing action $a'$ is at
least as likely to result in outcome $\varphi$ as performing action
$a$. 

Intuitively, this $\delta^\E_{a,a',\varphi}$ term ought to play a significant 
role in how we define blameworthiness: if $\ag$ does not believe that
he can have any effect on outcome $\varphi$, then we can hardly blame
him for its occurrence.  At the same time, though, this does not seem
to tell the whole story.  For if $\ag$ believed he could change the
outcome but only by giving up his life, we would not blame him for
$\varphi$'s occurence.  Thus there seems to be a second factor at
play, the expected cost $c(a)$ that $\ag$ ascribes to each action $a$.
Intuitively, the cost measures such factors as the cognitive 
effort, the time required to perform the action, the emotional 
cost of the action, and the potential negative consequences of
performing the action (like death).
(HK provide further discussion and intuition for cost.)

Noting that the balance between these two terms seems to be
situation-dependent, HK propose that a
parameter $N > \max_{a'}c(a')$ be used to weight the cost term in a
given scenario.  The degree of blameworthiness of $\ag$ for $\varphi$
relative to action $a'$, given that $\ag$ took action $a$, is then
defined to be $\db^{c}_{N}(a,a',\E,\varphi) = \delta^\E_{a,a',\varphi} \frac{N
	- \max(c(a') - c(a),0)}{N}$.  The degree of blameworthiness of $\ag$
for $\varphi$ given that $\ag$ took action $a$ is then $\db^{c}_{N}(a,\E,\varphi)
= \max_{a'} \db^{c}_{N}(a,a',\E,\varphi)$. 

As pointed out by HK,
blameworthiness judgments are not always made
relative to the beliefs of the agent.  It may be more appropriate to
consider the
beliefs that we believe that the agent \emph{ought} to have had.  Consider, for
example, a drunk driver who gets into an accident; in his inebriated
state, he may have believed that it was perfectly safe to drive, but we
still consider him blameworthy because we do not consider that belief
acceptable.
The definition just takes an epistemic state as input, without
worrying about whose epistemic state it is.

\subsection{Blameworthiness of groups}\label{sec:groupblame}

As HK already note, this definition of blameworthiness seems to
provide unsatisfactory results in settings where multiple agents are
involved.  Consider for instance the well-known \textit{Tragedy of the
Commons}~\cite{Hardin68}. 

\begin{example}\label{commons}
{\rm
$100$ fishermen live by a lake.  If at least $10$ of them
overfish this year then the entire fish population of the lake will
die out and there will be nothing left to fish in coming years.
Each fisherman, however, believes that it is very likely that at least $10$ other fishermen will overfish.  So given that all the fish will die out no matter his particular action, each fisherman decides to maximize his utility that year and so overfishes.  By the end of the year the entire fish population has died out.

Under the definition of blameworthiness discussed, each fisherman will
have blameworthiness close to $0$, as $\delta^\E_{a,a',\varphi}$ will be
close to $0$: each fisherman deemed the probability of the fish
population dying to be close to $1$ independent of his or her own
action.  And it seems unreasonable to say their beliefs were
unacceptable, as in fact what each fisherman predicted is exactly what
occurred.  But it seems problematic for each
fisherman to have a degree of blameworthiness that is almost 0
when the fishermen as a
group are clearly to blame for this outcome. 	  
}\wbox
\end{example}

How blameworthy are the fishermen for the outcome?  We claim
that in order to assign blame to the group, we need to assess the cost
to the fishermen of coordinating their actions, just as was done with
actions in the case of assigning individual blameworthiness.   
If in fact it was
impossible or extremely difficult for the fisherman to coordinate
(perhaps they had no means of communication, or all spoke different
languages), then they each should have very little
blameworthiness.
On the other hand, if coordination would have been relatively straightforward,
then the group should be viewed as quite blameworthy.

Computing the costs and expected effects of different ways that the group
could coordinate seems, in general, quite difficult.
To do this, we would need to have a model of what kinds of actions the
group could perform in order to bring about coordination.  Perhaps
some members of the group could convince politicians to pass laws that
put caps on how large the catch could be; perhaps they could arrange
for sensors that would be able to monitor how much each individual
fisherman caught.  Note that in this discussion we are not considering
whether fishermen would want to undertake these actions; only whether
there are feasible actions that might lead to coordination, and how
much they would cost.  
For example, the fishermen might choose not to lobby politicians to
pass laws because doing so would be quite expensive, but if it was
possible for them to do so, we still consider it an 
action the group could have taken. 
Unfortunately, while we may
understand how concrete actions might affect the likelihood of the
fish population dying out,
completely describing a set of rich causal models that capture the
possible dynamics that can lead to collaboration may be prohibitively
difficult, if not impossible. 

We instead consider a simpler way to capture difficulty of
coordination in group settings by abstracting away from these details.
We directly associate a cost with various distributions over
causal settings (i.e., epistemic states) that we view as the possible
outcomes of attempts to coordinate.  Intuitively, these are the
distributions that could have
been induced by feasible collective actions given beliefs about the probability of different causal settings in the richer models.
The cost of
a distribution represents the expected cost of performing whatever collective
actions were needed in the richer models to bring about that
distribution, as well as the expected costs of whatever actions will
be taken in the simpler model.  For example, suppose that the richer
causal model for 
the tragedy of the commons example allowed for installing sensors, and
after installing sensors we believe each fisherman $i$ will have an
independent probability $p_i$ of overfishing.
This leads to a distribution over the
settings of the simple causal models that were implicit in the description
of Example~\ref{commons}, where there is presumably an exogenous
variable $U_i$ that determines how much fishing fisherman $i$ does.
This exogenous variable is endogenous in the richer model, and is
affected by the installation of sensors.  In any case, the cost of that
distribution on causal settings in the simpler model is the cost of
performing whatever collective actions are required in the richer
model to arrange for the installation of sensors, plus the expected
collective costs of all the fishing the fishermen do.  Note that there
may 
not be feasible collective actions (i.e., ones with finite cost) in
the richer model that lead to none of the fishermen overfishing.

It is also worth noting that modeling the effects of an attempt at
coordination in the richer models as a distribution 
over causal settings in the simpler models allows us to capture other
effects that the coordination process may have.  For instance, 
consider a scenario where one of the fishermen
believes that taking everyone  
out on a boat ride on the lake would be a particularly effective way
to get the fishermen to feel social responsibility to not overfish.
Such a boat ride may also effect the pollution levels in the lake,
which in turn may also play a role in determining whether the fish 
population will die out.  By viewing the coordination process as
inducing a new distribution over causal settings, we can at the same
time capture both effects that the boat ride is expected to have on how
people behave and on pollution levels. 

With this background, we can now give analogues to the HK definitions.
We first give an analogue to the definition of
$\delta^\E_{a,a',\varphi}$.  As discussed above, rather than
comparing two actions that an individual can perform, we are comparing
two epistemic states (intuitively, ones brought about by different
collective actions in the richer model).
Let $\mathcal{E}_i = (Pr_i,\mathcal{K}_i)$, $i=1,2$, be two epistemic states.
Then, given an outcome $\varphi$, we can define the
extent to which $\varphi$ was more likely given $\E_1$ than $\E_2$ as 
\[ \delta_{\E_1,\E_2,\varphi}
= \max(0,{\Pr}_{1}(\intension{ \varphi}{\mathcal{K}_1}) -
{\Pr}_{2}(\intension{ \varphi }{\mathcal{K}_2})). \]
Just as in the HK definition, we are comparing the likelihood of
outcome $\varphi$ in two scenarios. For HK, the two scenarios were
determined by the agent performing two different actions; here, they
are determined by two different epistemic states that we can think of
 as arising from two coordination actions in the richer models combined with uncertainty regarding the richer causal setting.

We now want to get an analogue of the degree of blameworthiness
function $\db$ for a group.  Again, this will depend on two
parameters, a cost function $c$ and a parameter $N$ that determines
the relative weight we ascribe to the cost and the difference $\delta$
defined above.  However, now $c$ has different arguments.  One of its
arguments is, as suggested above, an epistemic state.  The second
is a subset of agents.
Let $\Ag = \{\ag_1, \dots, \ag_M \}$ be the set of all agents and
consider a
subset $\Ag' \subseteq \Ag$.
We think of $c(\Ag',\E)$ as the expected cost of the coordination actions
in the richer game required for the agents in $\Ag'$ to bring about
epistemic state
$\E$, plus the expected total costs of the actions of the agents in
the simpler models given that epistemic state (see below).%
\footnote{If there is more than one way for the agents in $\Ag'$ to
bring about $\E$, 
we can think of $c(\Ag',\E)$ as being the cost of the cheapest way to do so.}
The key point here is that we may not require coordination among all
the agents to bring about a particular epistemic state; it may only
require a subset.  Moreover different subsets of agents may have
different costs for obtaining the same outcome; the cost function is
intended to capture that.

The cost function is meant to take
into account not only the costs of bringing about $\E$, but the
expected cost of performing action $A_i$ for $i \in \Ag'$ in
the simpler causal models.  
This cost may vary from one causal model to another (e.g., it may be 
more costly to overfish if the probability of getting caught is
higher, and this may depend on the causal model).
Given an epistemic state,
we can compute the expected costs of performing $A_i$. 

Given a cost function $c$ and ``balance parameter'' $N$,
define the degree of blameworthiness for outcome
$\varphi$ of group $\Ag'$ and  epistemic state
$\mathcal{E}_1 = ({\Pr}_1,\mathcal{K}_1)$ relative to 
epistemic state $\mathcal{E}_2 = ({\Pr}_2,\mathcal{K}_2)$ such that
$c(\Ag',\E_2)$ is finite as
\[\begin{array}{ll}
&\gb^{c}_{N}(\Ag',\mathcal{E}_1,\mathcal{E}_2,\varphi)  \\
= &\delta_{\E_1,\E_2,\varphi} \frac{N - \max(c(\Ag',\mathcal{E}_2) -
	c(\Ag',\mathcal{E}_1),0)}{N}.
\end{array}
\]
Although we replace actions $a_1$ and $a_2$ in the definition of $\db$ by
epistemic states $\E_1$ and $\E_2$, and use a cost function with
different arguments,
the intuition for both the group
degree of blameworthiness function $\gb$ and the individual degree of
blameworthiness function $\db$ defined by HK are very much the same.

Just as for individual degree of blameworthiness, we can
define the group blameworthiness of group $\Ag'$ for outcome $\varphi$
given epistemic state $\mathcal{E}_1$ as the max over all possible
choices of $\E_2$. 
\[
\begin{array}{ll}
&\gb^{c}_{N}(\Ag',\mathcal{E}_1,\varphi)\\
= &\max_{\{\mathcal{E}_2 : c(\Ag',\E_2)\; \mathrm{is\
finite}\}} gb^{c}_{N}(\Ag',\mathcal{E}_1,\mathcal{E}_2,\varphi).
\end{array}
\]

Note that the degree of group blameworthiness of the empty group or any other
group that cannot coordinate any alternative actions will be $0$; the
only epistemic state $\mathcal{E}_2$ such that $c(\Ag',\E_2)$ is
finite will be $\mathcal{E}_1$ itself, and
$\delta_{\E_1,\E_1,\varphi}$ is $0$. 

One thing worth mentioning is that we require a monotonicity 
property for group blame: if
$\Ag'' \subseteq \Ag' \subseteq \Ag$ then $gb^{c}_{N}(\Ag'',\mathcal{E},\varphi) \leq
gb^{c}_{N}(\Ag',\mathcal{E},\varphi)$.  
The reason for this is that if group $\Ag''$
could coordinate in a particular way for a particular cost then that
subset of group $\Ag'$ could do exactly the same thing.  Essentially, when
considering the possibility of a group coordinating, we must really
consider the possibility of coordination of
any subset of that group.

Up to now, we have not said anything about \emph{whose} epistemic
state we should use for the epistemic states $\E_1$ and $\E_2$ in the definitions above.
In the case of individual blameworthiness, the typical
assumption is that they represent the epistemic state of the agent
whose blameworthiness is being considered, although as HK already
observed, it may at times be reasonable to assume that it is the
epistemic state that society thinks that agent should have.  Here we
are talking about group blameworthiness, so it is less clear whose
epistemic state should be used.  It is certainly not clear what a
``group epistemic state'' should be.  It still makes sense to think
about ``society's epistemic state''; that is, society's view of
what a reasonable agent's beliefs should be.  We can also take the
epistemic state to be the subjective beliefs of one of the agents.  Indeed,
we will often consider the epistemic state of an agent in the group.
We could also view the cost function as subjective---again, it could be
society's cost function or the cost function from the perspective of a
particular agent.
The definition is agnostic as to where the
epistemic state and cost function are coming from, but to apply the
definition we need to be explicit.

It is now worth returning briefly to Example~\ref{commons},
to see how this definition plays out there.
Given an epistemic state $\mathcal{E}_i =
(Pr_i,\mathcal{K}_i)$ and cost function $c_i$ representing the beliefs
of agent $\ag_i$, first consider a scenario where it would be
essentially impossible for the 
fishermen to coordinate (e.g., no two fishermen speak the same
language).
If $\ag_i$ believed this, then the cost of coordinating any
possible alternative distribution would likely be very
high, so the term $\frac{N - \max(c_i(\Ag',\E_2) - c_i(\Ag',\E_1),0)}{N}$
would be close to $0$.
Because this is true for all epistemic states $\E_2$, maximizing over 
$\E_2$ would still give that
$\gb^{c_i}_{N}(\Ag',\mathcal{E}_i,\varphi)$ is close to $0$.
On the other hand, suppose that $\ag_i$ believed that there was some
possible coordination of group $\Ag'$ that was not tremendously 
expensive and that could lead to an epistemic state $\E_2$ relative to which the probability of the fish population dying was lower
(e.g., imposing a fine on anyone who overfished). 
In this case,
to the extent that $\E_2$ was believed to be effective and low
cost, the group $\Ag'$ of fishermen would in fact be quite
blameworthy.
Note that $\Ag'$ might not consist of all the fisherman; it is
possible that a subset of fishermen is powerful enough to impose
fines.  In general, different subgroups will have different degrees
of blameworthiness.

\subsection{Apportioning group blameworthiness among agents}

Now that we have defined group blameworthiness, the question naturally
arises: how should group blameworthiness be apportioned among the
members of the group?
In this subsection, we suggest three axioms that we believe
apportionment of blame should satisfy.  It turns out that these axioms
are natural analogues of axioms that have been used to characterize
the Shapley value.
Shapley \shortcite{Shapley1953} introduced the Shapley value as an
approach to distributing benefits to individual agents in scenarios
where agents might coordinate to obtain greater total benefits than
they could individually.  The Shapley value has since also been
interpreted as way of appropriately distributing costs for shared
resources (see e.g.~\cite{RV79}).  It is thus not surprising that it
can  be applied in our setting as a way of apportioning group blame.

Given
a cost function $c$ and balance parameter $N$, let
$\db^{c,\mathcal{E}}_{N}(j,\varphi)$ be the 
degree of blameworthiness ascribed to $\ag_j$ for outcome $\varphi$ relative to
epistemic state $\E$. 
Consider the following three axioms for $\db^{c,\mathcal{E}}_{N}(j,\varphi)$: 
\begin{description}
	\item[\textit{Efficiency.}] 
All of the blame assigned to the full group of agents must be apportioned to the
agents in the group:
	\[
	 \displaystyle\sum_j \db^{c,\mathcal{E}}_{N}(j,\varphi) =
gb^{c}_{N}(\Ag,\mathcal{E},\varphi).
	\]

This axiom essentially encapsulates what we are trying to do:
apportion the total group blame among individuals.  Note that we do not
want an analogue of this for subgroups $\Ag'$ of $\Ag$.  For example,
if a small subgroup $\Ag'$ of fishermen cannot coordinate so as to affect the
outcome, they would have quite a low degree of blameworthiness,
although the group consisting of all the fishermen might have degree
of blameworthiness 1.  Thus, we do not necessarily want the sum of the 
degrees of blameworthiness of the individual fishermen in $\Ag'$ to be
the group blameworthiness of $\Ag'$.

\item[\textit{Symmetry}.] The names of agents should not
	affect their blameworthiness, so if we simply rename them then
	the blameworthiness ascribed to them should remain the same.
Formally, let $\pi$ be a permutation of $\{1,\dots,M\}$.
Given a model $M = ((\U,\V,\R,\G),\F)$, let $\pi \circ M
	= ((\U,\V,\R,\G'),\F)$, where $\G'(A) = \pi(\G(A))$ for all
	action variables $A \in \A$.  Given a set $\mathcal{K}$ of
causal settings, define $\pi \circ \mathcal{K} = \{ (\pi \circ M, \vec{u})
: (M,\vec{u}) \in \mathcal{K} \}$.
	That is to say, for any action that is assigned to agent $i$ in any model, we now instead assign it to agent $\pi(i)$.
	Given a distribution $\Pr$
over causal settings $(M,\vec{u})$, define $(\pi \circ \Pr)
	((\pi \circ M, \vec{u})) = \Pr((M,\vec{u}))$; if a setting had a particular probability then we want the corresponding setting with the actions renamed according to $\pi$ to have the same probability.
	Finally, given an epistemic state $\E = (\Pr,\K)$, let $\pi \circ \E = (\pi \circ \Pr, \pi \circ \K)$.
	The \textit{symmetry} axiom requires that   
	\[ 
\db^{c,\E}_{N}(i,\varphi) = \db^{\pi \circ c, \pi \circ
\E}_{N}(\pi(i),\varphi), 
	\]
where $(\pi \circ c)(\Ag',\E) = c(\{ b': \pi(b')\in \Ag' \},
\pi^{-1} \circ \E)$ (i.e., costs in the new models correspond to
costs pre-renaming, which we get by taking the $\pi$-preimage).  
	\item[\textit{Strong Monotonicity.}] If agent $\ag_j$
        contributes more to the group blameworthiness of all groups in
one scenario than another, then $\ag_j$ also ought to have a
        greater degree of (personal) blameworthiness in the first scenario.
        Formally, define the \emph{marginal contribution} of $\ag_j$ to
        the degree of blameworthiness of group $\Ag'$ as  
$$\begin{array}{ll}
\hspace{-.04in} \mb^{c,\mathcal{E}}_{N}(j,\Ag',\varphi) =\\
\hspace{-.04in} \left\{\begin{array}{lll}
\hspace{-.08in} \gb^{c}_{N}(\Ag',\mathcal{E},\varphi) -
		\gb^{c}_{N}(\Ag'\backslash\ag_j,\mathcal{E},\varphi)
&\hspace{-.12in} \mbox{if $\ag_j \!\in \! \Ag'$}\\
\hspace{-.08in} \gb^{c}_{N}(\Ag' \!\cup \! \ag_j,\mathcal{E},\varphi) \!-\!
\gb^{c}_{N}(\Ag',\mathcal{E},\varphi) &\hspace{-.12in} \mbox{if
$\ag_j \!\notin\! \Ag'.$}
\end{array}\right.
\end{array}
$$
Let $\mb^{c,\mathcal{E}}_{N}$ and $\mb'^{c,\mathcal{E}}_{N}$ be
the marginal contributions to the degree of blameworthiness for
        two different scenarios; let
        $\db^{c,\mathcal{E}}_{N} \textrm{ and
        } \db'^{c,\mathcal{E}}_{N}$ be the associated degree of
        (personal) blameworthiness for the two scenarios. 
	Then we require that if
	\[
	  \mb^{c,\mathcal{E}}_{N}(j,\Ag',\varphi) \geq \mb'^{c,\mathcal{E}}_{N}(j,\Ag',\varphi)\textrm{ for all } \Ag'\subseteq{\Ag}
	\]
	then
	\[
	 \db^{c,\mathcal{E}}_{N}(j,\varphi) \geq
	 \db'^{c,\mathcal{E}}_{N}(j,\varphi).
	\]
\end{description}

Young \shortcite{Young1985} showed that the only
distribution procedure that would satisfy Efficiency, Symmetry, and
Strong Monotonicity is the \emph{Shapley value}.
The Shapley value has an elegant closed-form expression.
It follows that the only way of assigning individual degree of
blameworthiness, given a group blameworthiness function $\gb$ has the form: 
\[
\begin{array}{ll}
\db^{c,\mathcal{E}}_{N}(j,\varphi) = \\
\sum\limits_{\{\Ag'\subseteq{\Ag} :\, \ag_j \in
        \Ag'\}} \frac{(|\Ag'|-1)! (|\Ag|-|\Ag'|)!}{|\Ag|!}
        mb^{c,\mathcal{E}}_{N}(j,\Ag',\varphi).
        \end{array}
\]

We thus have a technique for assigning a degree of blameworthiness for
an outcome to individuals in group settings.  However, this is
relative to an epistemic state, a cost function, and a balance
parameter.  The question still remains how these inputs should be
chosen.  As in HK, one approach when assigning a degree of
blameworthiness to an individual would be to take that individual's
epistemic state, cost function, and balance parameter.  But society may
decide that other choices are more reasonable.

Recall that in the last subsection we required that group blame always
be monotonic in the group, as if $\Ag''$ could coordinate in some
manner then they should also be able to do so as a subset of $\Ag'$.
It is not hard to see (and we show in the full paper) that
this suffices to ensure that individual
blameworthiness will always be non-negative.

Note that in the single-agent setting, where the
only agent choosing an action is some particular $\ag_1$,
if we assume (as HK implicitly did) that an agent can completely
decide his or her 
own actions without the decision process itself incurring costs beyond
the costs of the action, then 
the definition above agrees with the HK definition.
Consider $\ag_1$'s blameworthiness relative to $\ag_1$'s
epistemic state $\mathcal{E}_1 = (Pr_1, \mathcal{K}_1)$ and cost
function $c_1$.
Note that $|\Ag|=1$ and the only set $\Ag'$ containing $\ag_1$ is
$\{\ag_1\}$.  So 
there is only one term to sum over in 
$\db^{c_1,\mathcal{E}_1}_{N}(1,\varphi)$, and in that  
term we have that $\frac{(|\Ag|-1)!(|\Ag|-|\Ag|)!}{|\Ag|!} = 1$.  Thus, 
$\db^{c_1,\mathcal{E}_1}_{N}(1,\varphi) =
\mb^{c_1,\mathcal{E}_1}_{N}(1,\{\ag_1\},\varphi)$.  Because  
$\ag_1 \in \{\ag_1\}$, 
$\mb^{c_1,\mathcal{E}_1}_{N}(1,\varphi) =  
\gb^{c_1}_{N}(\{\ag_1\},\mathcal{E}_1,\varphi)
- \gb^{c_1}_{N}(\emptyset,\mathcal{E}_1,\varphi) =  
\gb^{c_1}_{N}(\{\ag_1\},\mathcal{E}_1,\varphi)$.
But now, because we assumed that an agent can decide his or
her own actions, the alternatives that group $\{\ag_1\}$ could have
coordinated will be precisely the set of actions available to $\ag_1$
at the costs they would incur to $\ag_1$, so this is the HK definition
of blameworthiness.

\subsection{An illustrative example}

The following example illustrates some features of these definitions: 
Consider a scenario where a committee of $7$ people, $\ag_1$ through
$\ag_7$, vote for whether or not to pass a bill.  If at least $4$
agents vote \yes, then the bill will pass.  Everyone agrees that it
would be better for the bill to pass, but there are external reasons
(such as opinions of constituents) that might result in agents
benefiting from voting \no\ as long as the  bill is passed.  The committee votes
and agents $\ag_1$ through $\ag_5$ all vote \no, so the bill does not
pass.  How blameworthy is each agent for this outcome? 

We now consider the degree of blameworthiness of some of the agents
and show how the degree of blameworthiness varies as a function of 
the agents' beliefs:
\begin{itemize}
	\item $\ag_1:$ $\ag_1$ believed that each of the $6$ other
	agents started with a $60\%$ chance of voting \yes.  For any
	coalition of $n$ agents, $\ag_1$ also believed that for a cost
of $n \times 100$ each agent's probability of voting \yes\
(including that of agents not in the coalition) could be
increased by $n \times 5 \%$ by applying social pressure.  In
addition, if $\ag_1$ herself was in the coalition, then for an
	additional cost of $2000$ she would have switched her vote
to \yes.
Given these beliefs, the degree of blameworthiness for the entire
group is $\approx0.390$, while $\ag_1$'s degree of blameworthiness is
$\approx 0.073$. 
\item $\ag_2:$ $\ag_2$ held essentially the same beliefs as $ag_1$,
	except that the additional cost necessary for her to change
her vote for coalitions that she was in was $500$
instead of $2000$.
Given these beliefs, the degree of blameworthiness of the entire
	group is $\approx0.390$, while $\ag_2$'s degree of blameworthiness 
is $\approx0.120$.

$\ag_2$ holds essentially the same beliefs as $\ag_1$, but her
	cost of changing her own vote to \yes\ is lower than
$\ag_1$'s.  The degree of blameworthiness of
	the entire group is the same with respect to both $\ag_1$'s
and $\ag_2$'s cost functions.
In the epistemic state that both agents share,
with $\ag_i$'s cost function (for $i=1,2$), 
the action
that maximizes degree of blameworthiness is 
the action where social pressure is applied by all, but
$\ag_i$ does not change her view.  Thus, the cost of
$\ag_i$ changing her view does not play
a role in determining the degree of group blameworthiness.  
	However, the blameworthiness of $\ag_2$ (according to
        $\ag_2$'s cost function) is
in fact greater than that of $\ag_1$ (according to $\ag_1$'s cost
	function), as
for some smaller groups the action that maximizes blame 
consists of $\ag_i$ changing her view, so the lower cost for
$\ag_2$ to do so will end up making her more blameworthy.  
This is
what we would expect; since it is less costly for $\ag_2$ to
	vote \yes, she intuitively ought to be more blameworthy for
	not doing so.
\item $\ag_3$ and $\ag_4$: $\ag_3$ held essentially the same
	beliefs as 
	$\ag_1$ except that she believed that social pressure
would be less effective.  In particular,
	she believed that a coalition of $n$ agents applying social
pressure for a cost of $n \times 100$ would result in
an increase of only $n \times 3 \%$ in each agent's probability of
voting \yes.
With these beliefs and cost function, the degree of blameworthiness
for the entire group 
is $\approx0.317$, and $\ag_3$'s degree of blameworthiness is
$\approx0.079$.

	$\ag_4$ held essentially the same beliefs as
$\ag_1$ except that she believed
that it would cost $n \times 150$ to get the social pressure applied
by $n$ agents to increase each agent's probability of
voting \yes\  by $n \times 5 \%$.
Given these beliefs, the degree of blameworthiness for the entire
	group is $\approx0.361$, and $\ag_4$'s degree of blameworthiness is
$\approx0.068$.

$\ag_3$ and $\ag_4$ each share beliefs similar to $\ag_1$'s, but
	they believe that social pressure will not be quite as
effective, either because it won't have as much of an impact or
	because it will be more costly.  As expected, in both of these
cases the degree of blameworthiness of the whole group decreases, as
	there is not as much 
	the group could have been expected to do to ensure that the
bill passed.  It is worth noting, however, that the
	blameworthiness of a particular agent may still go up, as it
	does here for $\ag_3$.  The reason for this is that, while the
	total group blame goes down, if the group does not have
effective alternatives to ensure the desired outcome, then it
may be even more important for that particular agent to take
an action that can significantly affect the outcome.  There are
	several factors that will affect whether (and to what extent)
	individual blameworthiness increases or decreases, such as the
	difference in cost, difference in expected effect, and the
	balance parameter $N$. 
	\item $\ag_5$: $\ag_5$ shared the same beliefs as $\ag_1$ with
	regard to what actions can be taken and the costs of taking
	those actions, but was more doubtful as to whether committee
	members would vote \yes\ without action being taken.  In
	particular, $\ag_5$ believed that each of the $6$ other agents
	started with a $40\%$ chance of voting \yes.  She still
	believed that a coalition of $n$ agents could increase each
agent's probability of voting \yes\ by $n \times 5 \%$ for a
price of $n \times 100$, and if she was in the coalition would
	have changed her vote to \yes\ for an additional cost of
$2000$.
With these beliefs and cost function, the degree of blameworthiness
for the entire 	group is $\approx0.560$, and $\ag_5$'s degree of
blameworthiness is $\approx0.125$.

	The only difference between $\ag_5$ and $\ag_1$ was that $\ag_5$ believed there was a higher probability of the bill failing in the first place. 
	Relative to this belief, $\ag_5$ (as well as the total group) is deemed 
	to be more blameworthy, as it is more critical that the group
	do something to ensure the bill have a higher change of
	passing.  To see how this plays out formally, consider the
	case where all 7 agents are involved in applying social
	pressure.  Then the effect this would have if the base
probability was $60\%$ would be a $\approx0.453$ increase in the
	probability of the bill passing.  If, on the other hand, the
	base probability was only $40\%$, then the social pressure
	would lead to a $\approx0.651$ increase in the probability of
a positive outcome.  It is worth noting that if the
	base probabilities of agents voting \yes\ were too low, then
	the blameworthiness would decrease, as the probability of the
	social pressure being able to actually effect a change would
	be low. 
	\item $\ag_6$: Finally, $\ag_6$ held exactly the same beliefs
and used the same cost function as $ag_1$, but unlike $ag_1$, she voted \yes.
In this case,  the degree of blameworthiness for the entire group
is  $\approx0.157$, and $\ag_6$'s degree of blameworthiness is
$\approx0.022$.

As we would expect, $\ag_6$ is deemed to be less blameworthy than
	$\ag_1$.  The total group blame is also lower relative to this
	epistemic state, as the probability of the bill failing to
pass is lower (because there is one
	definitive \yes\ vote) and so group action was less
	important. 
\end{itemize}

\section{Related Work}\label{sec:related}

Not surprisingly, there has been a tremendous amount of work on notions of blameworthiness across a wide range of fields.  In this section, we survey some of the literature most relevant to this work from computer science, philosophy, and law.

Our definitions of blameworthiness are based directly on those of HK.
Chockler and Halpern \shortcite{ChocklerH03} also defined a notion of
blame that is related to but somewhat different from blameworthiness;
see \cite{HKW18} for a discussion.

Our use of Shapley value in defining how to apportion group blame is
similar to (and partly inspired by) the work of Datta et
al. \shortcite{DDPZ15}. They define a measure of
the influence that each feature has on the classification of a
dataset.
So, for 
instance, if one feature is gender, their measure is intended to give
a sense of how much influence gender had on how the data was
classified.  They provide a set of desired axioms for influence and
show that there is a unique measure that satisfies these axioms, which
roughly corresponds to the probability that changing that feature
would change the classification.  This seems to have natural relevance
to our setting if we consider each feature to be the action of an
agent and the classification to be the outcome.  It is not sufficient,
however, as it is not clear how factors such as the cost of an action
(which is not relevant in the classification setting) should be
incorporated.
The Datta et al.~approach also does not deal with the ``group''
aspects of group blame.
It is in a sense closer to the work of HK than to ours.  
For example, in the Tragedy of the Commons, it would assign
a low degree of blameworthiness to individual agents.  While the group
aspects are not relevant in the setting of classification influence,
in our setting they are critical.  

Ferey and Dehez \shortcite{FD16} applied the Shapley value to 
sequential-liability tort cases, cases where the amount of damage each
agent's action brings about depends on the actions of earlier agents.
The court must decide how restitution of the damages should be
divided among the agents in such cases.
Ferey and Dehez used reasoning similar to ours to show that the Shapley value 
gives reasonable outcomes in this context. They also showed that the
outcomes seem to align well with some prior case law and legal literature.

There has been much work in the philosophy literature on
moral responsibility, including its nature and the conditions under
which one ought to be held morally responsible.  Particular
attention has been paid to the relation of moral responsibility to
such issues as free will and agency.
Eshleman \shortcite{Eshleman14} provides a good overview and further
references.  

There has also been significant discussion in the philosophical literature on issues of collective moral responsibility: can it ever really exist, under what conditions would it exist, can group moral responsibility be in turn divided among the member agents, and how ought it be divided if and when it can be?
May and Hoffman \shortcite{MH92} provide an
excellent collection of essays exploring some of the major ideas in
this area.
Cooper \shortcite{Cooper1968} argues that collective moral
responsibility is not always 
divisible among agents.  He considers an analogy of a delicious stew
made from various ingredients; we cannot say that any particular
ingredient has a specific degree of impact on the overall flavor;
rather, it is the precise way that the different flavors combined that led
to such a delicious stew.  Similarly, he argues, there may be
instances where no particular agent can be ascribed blame for the
mis-actions of the group, but rather it emerges from the collective as
a whole.
In these examples, it seems that Cooper would reject the Efficiency axiom.

Van de Poel et al.~\shortcite{PRZ15} focus on what they  call
  \emph{the problem of many hands} (a term originally due to
  Thompson \shortcite{Thompson80}): that is, the problem of allocating
  responsibility to individual agents who are members of a group that is clearly
    responsible for an outcome.  
They formalize some of their ideas using a variant
of the logic CEDL (\emph{coalition epistemic dynamic logic})
\cite{LR15}.
Unfortunately, CEDL cannot directly capture counterfactuals, nor can
it express quantitative notions like probability.  Thus, 
it cannot capture more quantitative
tradeoffs between choices that arise when defining degree of
blameworthiness.

Finally, it is worth mentioning some of the factors that come into
play in legal notions of blameworthiness.  Here we focus on two in
particular: \textit{joint and several liability}
and \textit{normality}.  In tort cases
where defendants are jointly and severally liable, each defendant can be 
considered to be independently liable for the full extent of damages.
Thus the injured party can recover the full amount of damages from
any of the defendants; it is up to that defendant who ends up paying
damages to then
attempt to recover some of the payment from other guilty
parties.  Thus, if two parties are guilty for an outcome but
one does not have the means to make restitution or is
inaccessible, the other party must make full restitution. 
This may be viewed as suggesting that there are
cases where the law deems each agent who is sufficiently responsible
as being fully blameworthy for the outcome rather than just having a
portion of the blameworthiness.
However, a more reasonable interpretation is that the law takes into
account considerations other than just degree of blameworthiness when
imposing penalties.  Nevertheless, considerations of blameworthiness
are likely to come into
play when the defendant who is compelled to pay attempts to recover
some damages from the other defendants.
When joint and several liability should be applied is a complicated
matter in the legal literature
(see, e.g., \cite{Prosser1936}).

Another notion at play in legal considerations of blameworthiness is
the legal norm.  The only considerations we have built into our
definitions are expected affect on the outcome and the cost of
actions.  In the law, however, the extent to
which an agent is judged to have deviated from the legal norm may play a
role in judgments of blameworthiness for outcomes that were brought
about by multiple individuals
\cite{ALI00}.
In future work we hope to further explore formalizations of some of
the notions at play in legal ascription of blameworthiness.
Work done on combining notions of normality with
causality \cite{Hal48,HH11} may prove relevant in dealing with issues
like legal norms.

\section{Conclusion}\label{sec:conclusion}

We have provided a way to ascribe blameworthiness to groups of agents
that generalizes the HK definition.
We then showed how, given ascriptions of
group blameworthiness, the 
Shapley value can be used to ascribe blameworthiness to individual
agents.
These two contributions are separable; if
an alternative definition of group blameworthiness is used,
the Shapley value could
still be used to ascribe blameworthiness to individual agents. 

In considering these issues carefully, one obvious question is whether
we view our definitions as descriptive or prescriptive.
The answer is ``both''.   We plan to do experiments to
see if the 
perceived difficulty of coordination really does affect how people
ascribe group blameworthiness, and to see whether an agent's potential
marginal contribution to an account affects his ascribed degree of
blameworthiness.   To the extent that we can view legal penalties as
proxies for degree of blameworthiness, we can also examine the legal
literature to see how these issues affected outcomes in legal cases
(although, as we observed earlier, there is clearly more to how
penalties are apportioned in legal cases than just blameworthiness).
Whether or not our definitions exactly match how people seem to
ascribe blameworthiness, we might still ask whether these definitions
might be useful as guides for ascribing blameworthiness in
situations involving  self-driving cars (or a combination of
self-driving cars and humans).

Formalizing notions of moral responsibility will be
critical for the eventual goal of designing autonomous agents that
behave in a moral manner.  We believe that blameworthiness as we have
considered it in this work is one important component of moral
responsibility, though not the whole story.  In future work we hope to
continue exploring how these notions can be formalized and applied to
a wide variety of settings, especially legal settings; we hope that
others will join us in  considering these problems. 

\paragraph{Acknowledgments:}
This work was supported in part by NSF grants IIS-1703846 and IIS-1718108,
ARO grant W911NF-17-1-0592, and a grant from the Open Philanthropy
project. We would like to thank Bruce Chapman for pointing
out the work of Ferey and Dehez, and the 
anonymous reviewers of the paper for comments that provided some
interesting food for thought. 

\bibliographystyle{aaai}
\bibliography{joe,z}

\begin{thebibliography}{}

\bibitem[\protect\citeauthoryear{{American Law Institute}}{2000}]{ALI00}
{American Law Institute}.
\newblock 2000.
\newblock {\em Restatement of the Law Third, Torts, Apportionment of
  Liability}.
\newblock American Law Institute Publishers.

\bibitem[\protect\citeauthoryear{Chockler and Halpern}{2004}]{ChocklerH03}
Chockler, H., and Halpern, J.~Y.
\newblock 2004.
\newblock Responsibility and blame: A structural-model approach.
\newblock {\em Journal of A.I. Research} 20:93--115.

\bibitem[\protect\citeauthoryear{Cooper}{1968}]{Cooper1968}
Cooper, D.~E.
\newblock 1968.
\newblock Collective responsibility.
\newblock {\em Philosophy} 43(165):258--268.

\bibitem[\protect\citeauthoryear{Datta \bgroup et al\mbox.\egroup
  }{2015}]{DDPZ15}
Datta, A.; Datta, A.; Procaccia, A.~D.; and Zick, Y.
\newblock 2015.
\newblock Influence in classification via cooperative game theory.
\newblock In {\em IJCAI},  511--517.

\bibitem[\protect\citeauthoryear{{De Lima} and Royakkers}{2015}]{LR15}
{De Lima}, T., and Royakkers, L. M.~M.
\newblock 2015.
\newblock A formalization of moral responsibility and the problem of many
  hands.
\newblock In Poel, I. v.~d.; Royakkers, L.; and Zwart, S.~D., eds., {\em Moral
  Responsibility and the Problem of Many Hands}. New York: Routledge.

\bibitem[\protect\citeauthoryear{Eshleman}{2016}]{Eshleman14}
Eshleman, A.
\newblock 2016.
\newblock Moral responsibility.
\newblock In Zalta, E.~N., ed., {\em The Stanford Encyclopedia of Philosophy
  {\rm (Winter 2016 edition)}}.
\newblock Available at
  http://plato.stanford.edu/archives/spr2014/entries/moral-responsibility/.

\bibitem[\protect\citeauthoryear{Ferey and Dehez}{2016}]{FD16}
Ferey, S., and Dehez, P.
\newblock 2016.
\newblock Multiple causation, apportionment, and the shapley value.
\newblock {\em The Journal of Legal Studies} 45(1):143--171.

\bibitem[\protect\citeauthoryear{Halpern and Hitchcock}{2015}]{HH11}
Halpern, J.~Y., and Hitchcock, C.
\newblock 2015.
\newblock Graded causation and defaults.
\newblock {\em British Journal for the Philosophy of Science} 66(2):413--457.

\bibitem[\protect\citeauthoryear{Halpern and Kleiman-Weiner}{2018}]{HKW18}
Halpern, J.~Y., and Kleiman-Weiner, M.
\newblock 2018.
\newblock Towards formal definitions of blameworthiness, intention, and moral
  responsibility.
\newblock In {\em Proceedings of the Thirty-Second AAAI Conference on
  Artificial Intelligence (AAAI-18)},  1853--1860.

\bibitem[\protect\citeauthoryear{Halpern and Pearl}{2005}]{HP01b}
Halpern, J.~Y., and Pearl, J.
\newblock 2005.
\newblock Causes and explanations: a structural-model approach. {P}art {I}:
  {C}auses.
\newblock {\em British Journal for Philosophy of Science} 56(4):843--887.

\bibitem[\protect\citeauthoryear{Halpern}{2016}]{Hal48}
Halpern, J.~Y.
\newblock 2016.
\newblock {\em Actual Causality}.
\newblock Cambridge, MA: MIT Press.

\bibitem[\protect\citeauthoryear{Hardin}{1968}]{Hardin68}
Hardin, G.
\newblock 1968.
\newblock The tragedy of the commons.
\newblock {\em Science} 162:1243--1248.

\bibitem[\protect\citeauthoryear{May and Hoffman}{1992}]{MH92}
May, L., and Hoffman, S.
\newblock 1992.
\newblock {\em Collective Responsibility: Five Decades of Debate in Theoretical
  and Applied Ethics}.
\newblock Rowman \& Littlefield Publishers.

\bibitem[\protect\citeauthoryear{Poel, Royakkers, and Zwart}{2015}]{PRZ15}
Poel, I. v.~d.; Royakkers, L.; and Zwart, S.~D.
\newblock 2015.
\newblock {\em Moral Responsibility and the Problem of Many Hands}.
\newblock New York: Routledge.

\bibitem[\protect\citeauthoryear{Prosser}{1936}]{Prosser1936}
Prosser, W.~L.
\newblock 1936.
\newblock Joint torts and several liability.
\newblock {\em California Law Review} 25:413.

\bibitem[\protect\citeauthoryear{Roth and Verrecchia}{1979}]{RV79}
Roth, A.~E., and Verrecchia, R.~E.
\newblock 1979.
\newblock The {S}hapley value as applied to cost allocation: a
  reinterpretation.
\newblock {\em Journal of Accounting Research}  295--303.

\bibitem[\protect\citeauthoryear{Shapley}{1953}]{Shapley1953}
Shapley, L.~S.
\newblock 1953.
\newblock A value for $n$-person games.
\newblock {\em Contributions to the Theory of Games} 2(28):307--317.

\bibitem[\protect\citeauthoryear{Thompson}{1980}]{Thompson80}
Thompson, D.~E.
\newblock 1980.
\newblock Moral responsibility and public officials: the problem of many hands.
\newblock {\em American Political Science Review} 44(3):905--916.

\bibitem[\protect\citeauthoryear{Young}{1985}]{Young1985}
Young, H.~P.
\newblock 1985.
\newblock Monotonic solutions of cooperative games.
\newblock {\em International Journal of Game Theory} 14(2):65--72.

\end{thebibliography}
\end{document}